\documentclass[%
reprint,
 amsmath,amssymb,
aip,
]{revtex4-1}

\usepackage{color}

\usepackage{graphicx}
\usepackage{dcolumn}
\usepackage{bm}

\begin{document}

\preprint{AIP/123-QED}

\title{Cavity-enhanced high harmonic generation for XUV  time-resolved ARPES}


\author{A.K. Mills}
\affiliation{Department of Physics and Astronomy, University of British Columbia, Vancouver, British Columbia V6T 1Z1, Canada}
\affiliation{Quantum Matter Institute, University of British Columbia, Vancouver, British Columbia V6T 1Z4, Canada}

\author{S. Zhdanovich}
\affiliation{Department of Physics and Astronomy, University of British Columbia, Vancouver, British Columbia V6T 1Z1, Canada}
\affiliation{Quantum Matter Institute, University of British Columbia, Vancouver, British Columbia V6T 1Z4, Canada}

\author{M.X. Na}
\affiliation{Department of Physics and Astronomy, University of British Columbia, Vancouver, British Columbia V6T 1Z1, Canada}
\affiliation{Quantum Matter Institute, University of British Columbia, Vancouver, British Columbia V6T 1Z4, Canada}

\author{F. Boschini}
\affiliation{Department of Physics and Astronomy, University of British Columbia, Vancouver, British Columbia V6T 1Z1, Canada}
\affiliation{Quantum Matter Institute, University of British Columbia, Vancouver, British Columbia V6T 1Z4, Canada}

\author{E. Razzoli}
\affiliation{Department of Physics and Astronomy, University of British Columbia, Vancouver, British Columbia V6T 1Z1, Canada}
\affiliation{Quantum Matter Institute, University of British Columbia, Vancouver, British Columbia V6T 1Z4, Canada}

\author{M. Michiardi}
\affiliation{Department of Physics and Astronomy, University of British Columbia, Vancouver, British Columbia V6T 1Z1, Canada}
\affiliation{Quantum Matter Institute, University of British Columbia, Vancouver, British Columbia V6T 1Z4, Canada}
\affiliation{Max Planck Institute for Chemical Physics of Solids, 01187 Dresden, Germany}

\author{A. Sheyerman}
\affiliation{Department of Physics and Astronomy, University of British Columbia, Vancouver, British Columbia V6T 1Z1, Canada}
\affiliation{Quantum Matter Institute, University of British Columbia, Vancouver, British Columbia V6T 1Z4, Canada}

\author{M. Schneider}
\affiliation{Department of Physics and Astronomy, University of British Columbia, Vancouver, British Columbia V6T 1Z1, Canada}
\affiliation{Quantum Matter Institute, University of British Columbia, Vancouver, British Columbia V6T 1Z4, Canada}

\author{T.J. Hammond}
\altaffiliation[Currently at ]{Department of Physics, University of Windsor,  Windsor, Ontario N9B 3P4, Canada}
\affiliation{Department of Physics and Astronomy, University of British Columbia, Vancouver, British Columbia V6T 1Z1, Canada}

\author{V. S\"uss}
\affiliation{Max Planck Institute for Chemical Physics of Solids, 01187 Dresden, Germany}

\author{C. Felser}
\affiliation{Max Planck Institute for Chemical Physics of Solids, 01187 Dresden, Germany}

\author{A. Damascelli}
\affiliation{Department of Physics and Astronomy, University of British Columbia, Vancouver, British Columbia V6T 1Z1, Canada}
\affiliation{Quantum Matter Institute, University of British Columbia, Vancouver, British Columbia V6T 1Z4, Canada}

\author{D.J. Jones}
\email{djjones@physics.ubc.ca}
\affiliation{Department of Physics and Astronomy, University of British Columbia, Vancouver, British Columbia V6T 1Z1, Canada}
\affiliation{Quantum Matter Institute, University of British Columbia, Vancouver, British Columbia V6T 1Z4, Canada}


\date{\today}

\begin{abstract}
With its direct correspondence to electronic structure, angle-resolved photoemission spectroscopy (ARPES) is a ubiquitous tool for the study of solids. When extended to the temporal domain, time-resolved (TR)-ARPES offers the potential to move beyond equilibrium properties, exploring both the unoccupied electronic structure as well as its dynamical response under ultrafast perturbation. Historically, ultrafast extreme ultraviolet (XUV) sources employing high-order harmonic generation (HHG) have required compromises that make it challenging to achieve a high energy resolution\textemdash which is highly desirable for many  TR-ARPES studies\textemdash while producing high photon energies and a high photon flux. We address this challenge by performing HHG inside a femtosecond enhancement cavity (fsEC), realizing a practical source for TR-ARPES that achieves a flux of over 10$^{11}$~photons/s delivered to the sample, operates over a range of 8-40~eV with a repetition rate of 60 MHz. This source enables TR-ARPES studies with a temporal and energy resolution of 190~fs and 22~meV, respectively. To characterize the system, we perform ARPES measurements of polycrystalline Au and MoTe$_2$, as well as TR-ARPES studies on graphite.
\end{abstract}

\maketitle
\section{Introduction} \label{sec:Intro}
Angle-resolved photoemission spectroscopy (ARPES) has emerged as a critically important technique to study the electronic structure of matter. The extension to time-resolved ARPES (or TR-ARPES) employs the usual pump-probe technique  to monitor the relaxation pathways of excited electrons with the goal of uncovering new insights into the equilibrium state itself or characteristic behaviors not observed in steady state\cite{Armitage2014}.  TR-ARPES provides the most complete picture of the excitation and intrinsic timescale(s) of the optically-driven system \cite{Smallwood2012a,Boschini2018,Cilento2018, Sobota2012,PefettiTI2012,Gierz2013,HofmannTRgraphene2013, PerfettiTaS2,BovensiepenTRGd,SentefPRX}, and consequently, it has been applied to study a wide range of quantum systems \cite{Gedik2017}, such as high-temperature superconductors \cite{Perfetti2007,Smallwood2012a,NatCommBovensiepen2016,DessauPRX2017, Boschini2018, Cilento2018}, charge-ordered compounds \cite{Schmitt1649_2008,Rohwer2011}, topological materials \cite{Sobota2012,PefettiTI2012} and 2-D materials \cite{Gierz2013, HofmannTRgraphene2013}.

Equilibrium ARPES has advanced significantly in its ability to map electronic structure with high resolution over a large range of energy and momentum space, benefitting from advances in electron-analyzer technology and quasi-continuous photon sources such as UV discharge lamps and third-generation synchrotrons. Likewise, TR-ARPES using ultrafast pulsed lasers and frequency conversion with nonlinear crystals has been applied to the study of quantum matter on short timescales (with accompanying low energy resolution)\cite{Faure2012a,Boschini2014} and also with very high energy resolution (and long timescales) \cite{Liu2008a,Okazaki2012a}. Ultrafast lasers can also be used for high-order harmonic generation\cite{McPherson:87,Ferray1988,Corkum1993} (HHG) to generate photons with energies spanning the vacuum ultraviolet (VUV), XUV, and soft x-rays \cite{Gibson2003}.  HHG sources have been applied to photoemission spectroscopy since the 1990s\cite{Haight1994a}, yet they have not seen the same widespread adoption in laboratory-based TR-ARPES as have nonlinear crystal-based UV sources, at least in part due to their overall complexity.  Furthermore, it remains a major technical challenge to build a single laser source ideally suited for a wide range of possible experiments due to the coupled and often contradictory requirements of TR-ARPES, including the photon energy, time and energy resolution, photon flux, and pump fluence.
\begin{figure*}[t]
\fbox{\includegraphics[width=2\columnwidth]{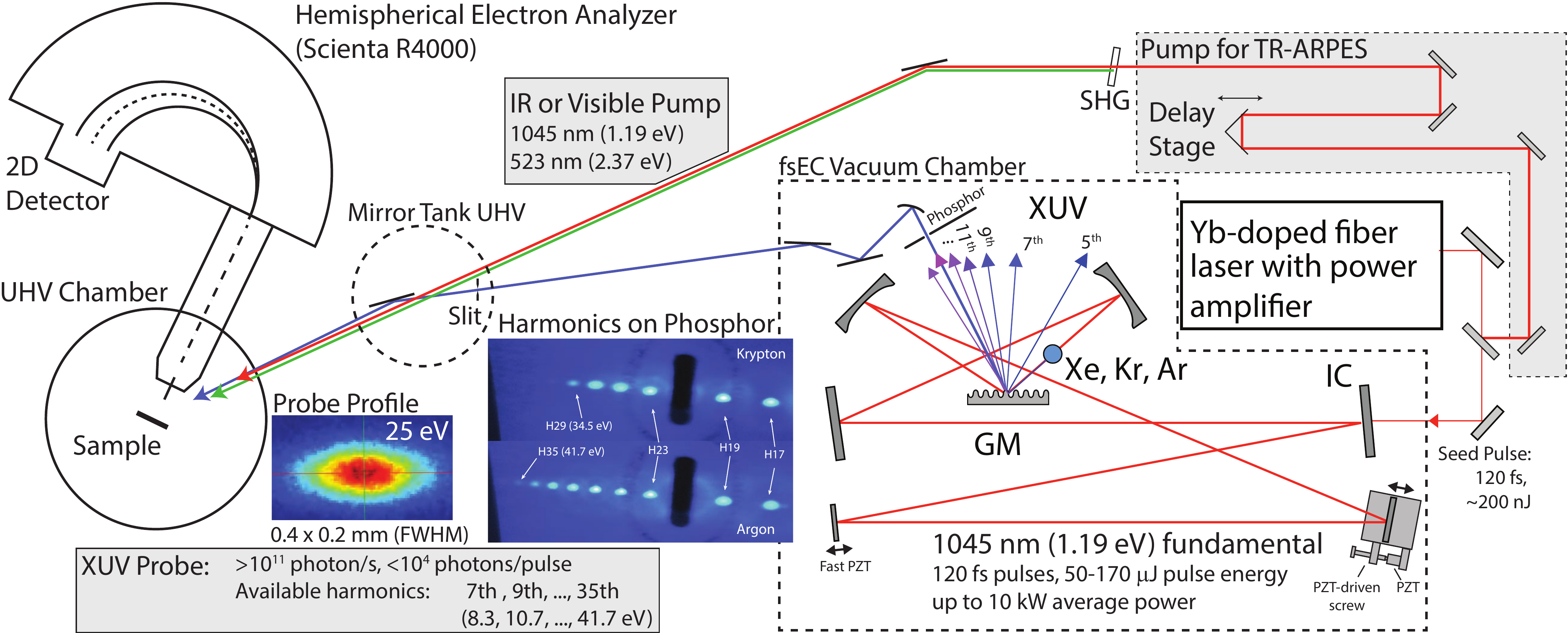}}
\caption{A custom femtosecond laser system operating at 1045 nm delivers a 60-MHz pulse train to the femtosecond enhancement cavity (fsEC). A portion of the pulse train is also employed for the pump beam for time- and angle-resolved photoemission spectroscopy (TR-ARPES). Harmonics out to about 40~eV (in argon) are produced in the fsEC, and are out-coupled and spectrally separated from the fsEC by the grating mirror (GM). In the results discussed here, H21 (25 eV) is collimated and then imaged on the sample in the electron analyzer at distance 2.5 m away, after passing through a three-stage differential pumping system. The XUV beam profile (in the plane normal to propagation direction) at the sample is shown in the heat map.}
\label{fig:setup}
\end{figure*}

Practically speaking, the probe for TR-ARPES must have a photon energy of at least 6~eV to exceed the work function of the material under investigation (4-5~eV for most materials). Since the momentum of the photoelectron is proportional to the square root of its kinetic energy, increasing the photon energy (and hence the kinetic energy of the photoelectron) beyond this minimum allows access to a larger range of momentum ({\it k}-) space within the Brillouin zone\cite{Damascelli2004}. Similar to a synchrotron source, the photon energy should also be tunable\textemdash up to 30 eV or above\textemdash to probe three-dimensional {\it k}-space and address photon energy dependent matrix element effects, which modify the spectral weight in ARPES as well as the overall photoemission cross-section. The time and energy resolution should be well-suited to the ultrafast processes being studied\cite{Giannetti2016}. For example, electron-electron interactions occur on timescales $\ll$100~fs, necessitating correspondingly short pulses. Electron-phonon scattering occurs on 100-fs time scales or greater, thus making it possible to apply TR-ARPES with an (experimentally significant) improved energy resolution of $<$50 meV, to probe the ultrafast dynamics of the electron-phonon interaction \cite{Na2018}.

Achieving sufficient signal statistics in ARPES typically requires long data acquisition times.  Space-charge distortions\cite{Damascelli, Hellmann2009} limit the number of photo-emitted electrons per pulse as well as the effective energy and momentum resolutions, and therefore one must increase count rates by increasing the pulse rate, not the number of photons per pulse. Increasing the repetition rate is beneficial for at least two additional reasons:  (i) to enable studies of samples that degrade rapidly even in ultrahigh vacuum (UHV), and (ii) to limit noise of technical origin ({\it e.g.} drift) that is difficult to control over long measurements.  Finally, the improved signal-to-noise ratio achieved with high-repetition rate TR-ARPES reduces the pump fluence required to detect a signal, and thus provides a route toward perturbative studies of quantum matter \cite{Rohwer2011,Boschini2018,Steinleitner2017}. The upper limit on the repetition rate is sample-dependent and related to the average power imparted by the pump field.  

Here we present an XUV source, shown schematically in Fig.\,\ref{fig:setup}, based on a femtosecond enhancement cavity (fsEC) that is extremely well-suited for perturbative TR-ARPES studies of quantum materials over the full Brillouin zone.  The source operates at a repetition rate of 60~MHz and produces a useable photon-energy tuning range of 8-40~eV.  For the 21st harmonic (H21) at a photon energy of 25~eV, it delivers over 10$^{11}$~photons/s to the sample with an energy resolution of 22~meV, and a temporal resolution of 190~fs.  It has been in service as a dedicated TR-ARPES source with several months of continuous operation. In the following, we discuss the design and construction of this source, and we present an experimental characterization of its properties, including long-term stability, time/energy resolution, and its reach in {\it k}-space. We demonstrate the potential of the source with a study of equilibrium ARPES on 2H-MoTe$_2$ over the full Brillouin zone, and of ultrafast dynamics of excited electrons in graphite at the edge of the Brillouin zone as measured by high-energy resolution TR-ARPES.

\section{The Amplified Yb-doped Fiber Laser}
\label{sec:Laser}
We use the output of an amplified Yb-doped fiber laser system operating at a repetition rate of 60~MHz to seed the fsEC.  The laser is comprised of a custom-built mode-locked Yb-doped fiber oscillator, a single-mode fiber (SMF) preamplifier, and a photonic crystal fiber (PCF) power amplifier.  The choice to build a custom laser system was made to provide more control of the resulting pulse characteristics, such as minimizing the noise on the carrier-envelope offset frequency (f$_{CEO}$) and maintaining good compressibility of the pulse at the end of the amplifier chain, both of which can change substantially with oscillator configuration.

\begin{figure*}[t]
\fbox{\includegraphics[width=2\columnwidth]{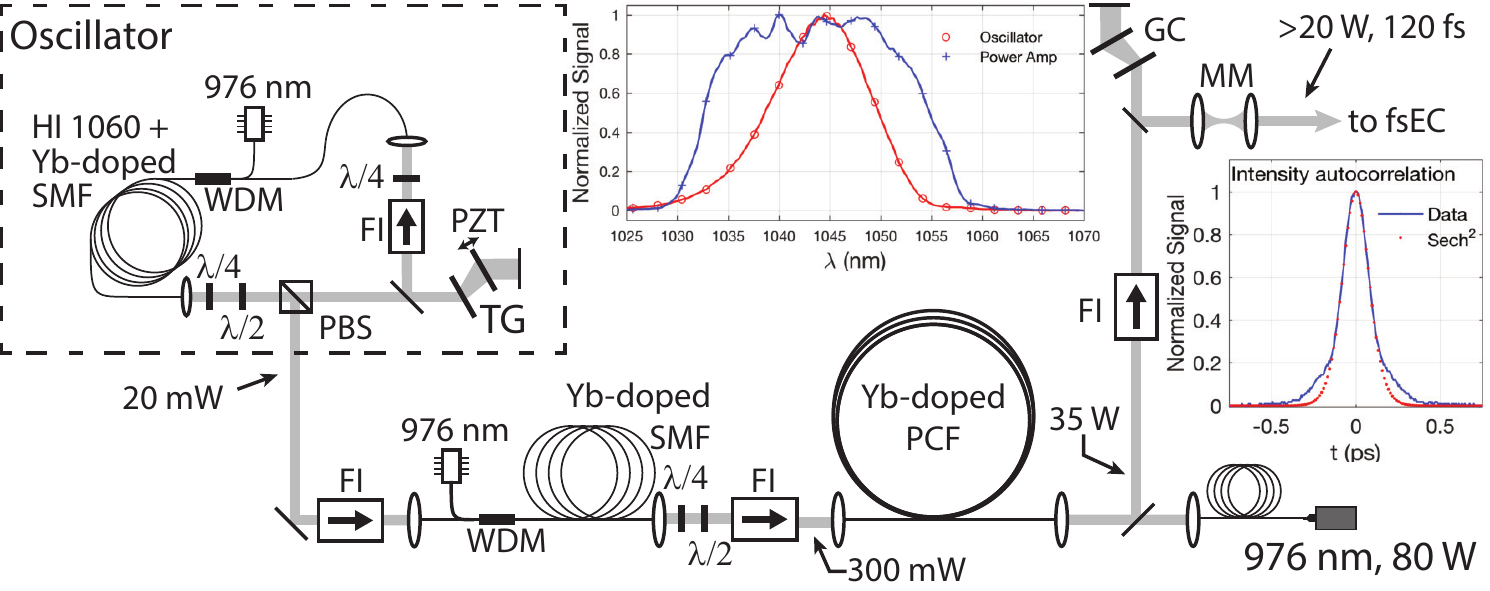}}
\caption{The laser system consists of a Yb-doped fiber oscillator that utilizes intracavity transmission gratings (TG) to control the dispersion in the oscillator as well control the carrier-envelope offset phase of the oscillator (via PZT-actuated grating separation).  The free-space output of the oscillator is directly coupled to a single-mode fiber (SMF) preamplifier that produces an output power of approximately 300 mW, which is coupled into the photonic crystal fiber (PCF) power amplifier.  The power amplifier produces up to 35 W of output power, which yields more than 20 W after a transmission grating compression (GC) stage. A mode matching (MM) telescope is used to match the laser mode to the TEM$_{00}$ mode of the fsEC.}
\label{fig:lasersystem}
\end{figure*}

The mode-locked oscillator is based on nonlinear polarization evolution (NPE)\cite{Nelson1997} and is comprised of an SMF section and a free-space section, as shown in Fig.\,\ref{fig:lasersystem}.  
We operate the oscillator in a regime where the net dispersion is slightly anomalous and the cavity dispersion limits the spectral bandwidth to 12~nm, centered at 1045~nm. The fiber section consists of two free-space collimators, approximately 30~cm of Yb-doped fiber, and a wavelength division multiplexer (WDM) to couple the 976~nm pump into the cavity.  The output of the oscillator is taken from the polarizing beamsplitter (PBS) and is approximately 20~mW.

The oscillator is robust and has been able to a maintain its spectral shape and pulse characteristics for nearly two years of continuous operation, without the need for realignment.  
Low noise on f$_{CEO}$\cite{Nugent-Glandorf2011} is important to reduce intensity noise in the fsEC, and we find our operating conditions sufficient to meet these requirements when the intensity noise of the oscillator is minimized\cite{Li2016a}.

The output of the oscillator is coupled into a preamplifier without any additional pulse stretcher.  The preamplifier consists of non-polarization maintaining, Yb-doped SMF (35~cm length), pumped with an SMF-coupled diode at 976~nm, delivering 600~mW. Nearly 300~mW is achieved at the output of the Faraday isolator (FI) placed after the preamplifier.  This beam is directed into the PCF amplifier, which is a 3~m length of Yb-doped PCF.  The PCF is pumped with up to 80~W at 976~nm, and we achieve an output power from the PCF as high as 35~W.  Safe operation at these power levels requires use of isolators before and after the PCF amplifier and an interlock system that turns off the pump diode current in the event that the input to the amplifier is somehow disrupted.  

The output of the PCF is compressed with a transmission grating compressor (GC), and a mode-matching (MM) telescope is used to produce the required input mode to couple to the fsEC.  For the work presented here, we operate with a PCF output power of 17~W, corresponding to 11~W delivered to the fsEC (losses are primarily due to pulse compression, and beam diagnostics).  The pulse duration after compression is typically 120~fs full width at half maximum (FWHM).

Finally, we implement an active beam stabilization system to improve the pointing stability of the laser coupled into the fsEC. This system reduces both the effective warm-up time of the system and the fsEC intensity noise by reducing coupling instabilities caused by environmental factors such as air condition, building vibrations, and thermal pointing drift of the PCF amplifier.

\section{The Femtosecond Enhancement Cavity}
\label{sec:fsEC}
The fsEC is a passive resonator in which ultrafast laser pulses are coherently combined to achieve a much more energetic pulse circulating in the resonator than is produced by the seed laser source. They are particularly useful in frequency conversion applications \cite{Jones2005,Gohle2005} with intrinsically low conversion efficiency (such as HHG), making enhancement factors of a few hundred or more readily achievable.  Early applications of fsECs as HHG sources were limited by numerous challenges to their long term operation\cite{Mills2012}, but development of applications with fsECs operating for extended periods has accellerated recently\cite{Corder2018a,Saule2018,Zhao2018a}.  In this work we report results from several months of intense use. We routinely operate the system for a week at a time, and we regularly observe lengthy `cavity-lock' periods as long as 16 hours.  Furthermore, much of the complexity related to operating the active stabilization system can be automated.  

Our fsEC (shown in Fig.\,\ref{fig:setup}) is a seven-mirror cavity, consisting of three highly reflective planar mirrors, a diffraction grating mirror (GM)\cite{Yost2008,Yost2011}, an input coupler (IC) with 0.45\% transmission, and two mirrors with a 15~cm radius of curvature.  The s-polarized cavity mode has a tight focus (1/e$^2$ radius of 20~$\mu m$) in between the two curved mirrors. To achieve resonant enhancement over the full spectrum of the incident laser pulses, the cavity round-trip time must be exactly matched to the laser pulse period\cite{Mills2012}\textemdash approximately 5~m at a repetition rate of 60~MHz. We achieve an enhancement factor of 300-500, depending how clean the GM is (as discussed in Sec.~\ref{sec:XUVSource}).

We have observed $>$10~kW of average power in the fsEC, although we typically run at lower powers for HHG operation with xenon and krypton gas. We do not actively cool the cavity mirrors, but we do observe a shift in the cavity length due to heating of the mirror substrates and the fsEC breadboard.  The rate of heating slows gradually over several hours, and reaches a steady state after about 24 hours of continuous operation.  An example of the locking performance is shown in Fig.\,\ref{fig:char}(d), where we demonstrate $<$0.5\% drift of the intracavity average power over a period of more than 16 hours during an ARPES measurement (note that the vertical scale of the intracavity fsEC signal is expanded by a factor of 10).


To achieve stable resonance enhancement of the laser pulses within the fsEC, two degrees of freedom must be actively stabilized: (i) the round-trip time of the pulse circulating in the fsEC relative to the laser pulse period, and (ii) the round-trip phase shift of the electric field carrier relative to the pulse envelope ({\it i.e.} the carrier-envelope phase, $\phi_{CEO}$).  In the frequency domain, these degrees of freedom correspond to the mode spacing of the fsEC and $f_{CEO}$ of the oscillator, respectively. We modify the fsEC round trip time (via the cavity length) to stabilize it relative to the oscillator repetition period ($1/f_{rep}$), and we adjust $\phi_{CEO}$ of the oscillator to match the fsEC round trip carrier-envelope phase shift.  This stabilization is performed with three servo control loops: (i) a fast, low dynamic range piezoelectric transducer (PZT) loop acting on the fsEC length, (ii) a high dynamic range, slow PZT loop that corrects for slow variations of the fsEC length, and (iii) a slow PZT loop controlling the spacing of the transmission gratings in the oscillator to change  $\phi_{CEO}$.
The control loops are implemented using a variation of the Pound-Drever-Hall (PDH) technique\cite{Black2001}. In our case, we apply a sinusoidal modulation (500-700~kHz) to the fast PZT, and we derive error signals for the cavity length and f$_{CEO}$ by sampling and demodulating photodetector signals at two different wavelengths in the spectrally dispersed fsEC reflection.  The output of the fast PZT servo is low-pass filtered and used as the error signal for the slow PZT servo.

Finally, we have implemented a microcontroller-based automatic locking system designed to allow both manual-mode operation of the servo system as well as an auto-mode to make the system user friendly. The autolock system is able to reacquire the system lock (typically in less than one second) in the event of a lock-breaking disturbance, and when necessary it corrects for long-term variations (such as thermal drift) that push the resonant cavity length beyond the dynamic range of the slow PZT.

\section{Intracavity HHG XUV Source}
\label{sec:XUVSource}

Early applications of fsEC HHG sources demonstrated great potential for high photon flux XUV at the high repetition rate preferred for applications such as frequency comb spectroscopy and ARPES. They also faced significant technical challenges, including intracavity plasma effects that limit the HHG efficiency and hydrocarbon contamination that limited long term operation\cite{Mills2012}.  Several approaches were established to mitigate the plasma effects on fsEC operation\cite{Carlson2011,Yost2011,Allison2011,Pupeza2013}. With improved high-power mirror coatings and in-situ ozone cleaning of fsEC mirrors, the community rapidly converged on solutions for the hydrocarbon contamination problem as well, and we demonstrated viable long-term operation for photoemission spectroscopy in 2014\cite{Mills2014,Mills2015}.

Optimizing HHG at the high repetition rate ($>$10~MHz) and low pulse energy ({\it e.g.} 100 $\mu J$) used in fsECs requires operation in the tight-focus regime, which is quite different from the widely studied operating regime of low-kHz repetition rate sources\cite{Gibson2003,Eich2014b,Rohde2016}.  On the other hand, the pulse energies used in fsECs are comparable to those used by state-of-the-art single-pass systems working in the 0.1-1~MHz range, enabled by advances in femtosecond laser technology \cite{Chen2009,Hadrich2010,Hadrich2011,Heyl2012a,Cabasse2012,Chiang2012,Wang2015}.  Motivated by the development of these sources, generalized power scaling laws for phase matching of HHG have been worked out\cite{Hadrich2016,Heyl2017}.  With respect to fsEC sources specifically, recent work to push the limits of HHG power scaling has focused on the effects of the plasma that persists in the fsEC focal region between pulses.  Working with shorter pulses is one way to reduce the total ionization\cite{Pupeza2013}, but not without increasing the spectral bandwidth of the resulting harmonics, which is undesirable for high-energy resolution ARPES.  At least two additional approaches are feasible: (i) increasing the velocity of the gas jet by increasing the temperature or mixing gases \cite{Porat2018}, and (ii) reducing the repetition rate and/or or laser mode size such that atoms in the gas jet interact with only one laser pulse \cite{Saule2018}.

In our system, HHG is performed in the tight focus of the fsEC, where a gas jet of xenon, krypton, or argon is delivered by a  quartz capillary with a 150~$\mu m$ diameter (backed with up to 1~ATM gas pressure). The generated harmonics are s-polarized and propagate collinearly with the fundamental, and are out-coupled by the GM, placed 3~cm after the fsEC focus. The GM has a period of 200~nm and a calculated diffraction efficiency of 2-4\% over the range of harmonics we produce.  The (blue) phosphor image in Fig.\,\ref{fig:setup} shows harmonics H17-H35 produced in krypton and argon (H21 passes through the slit in the plate).  During operation of the source, the harmonics in this image can be monitored in real-time (exposure times of a few ms) to monitor the power.  Figure~\ref{fig:char}(c) demonstrates a $\pm5\%$ drift in H23 and H25 during an ARPES measurement over a period of 16 hours.  No special care is taken to stabilize the gas pressure or gas jet alignment, suggesting improvement in XUV stability could be achieved with some effort.

We estimate that we generate $>10^{13}$ photons/s per harmonic in krypton, determined from the calculated diffraction efficiency of the GM and an XUV photodiode (with directly-deposited Al filter) measurement of the power in H21.  The manufacturer's quoted photodiode responsivity does not account for the Al$_2$O$_3$ layer on the filter, and thus represents a conservative estimate of the photon flux. For the measurements presented here, we first collimate H21 using a toroidal mirror with a 15~cm focal length, and then direct it to the ARPES chamber using three additional mirrors coated with boron carbide (combined reflectivity of all mirrors is calculated to be 20\%).  When the GM efficiency is included, we estimate that we deliver $>10^{11}$ photons/s to the sample.  The XUV beam can be imaged on a Ce:YAG scintillator crystal in both the mirror tank and ARPES vacuum chambers (see Fig.\,\ref{fig:setup}) to assist alignment though the differential pumping stages onto the sample.

To limit the degradation of the fsEC mirrors that occurs when they are exposed to XUV radiation \cite{Mills2012} we continuously flow a dilute mix of ozone (O$_3$) within oxygen (O$_2$) onto the GM and one mirror immediately after the GM, which helps us to maintain a reasonably high cavity enhancement. Only these two mirrors are directly flooded with O$_2$/O$_3$ as they experience the greatest XUV flux, though the chamber is sufficiently back-filled to provide some cleaning of other mirrors as well.  Over four months of near-continuous operation, we removed the GM and two other mirrors to clean them in an external plasma cleaner on one occasion.

It is worth noting that the GM output coupler is not a conventional choice for ultrafast experiments, as it is well known that diffraction gratings lead to a tilted pulse-front and a spatial chirp due to the angular dispersion of the grating \cite{WeinerUltrafastOptics}, thus increasing the effective pulse duration.  Time-preserving monochromators have been developed to separate harmonics while preserving femtosecond pulses using one or two gratings\cite{Poletto2004,Frassetto2011,Frassetto2014}. In our application, we weigh the consequences of the GM against its benefits.  In particular, the GM output coupler has a relatively high overall efficiency for out-coupling and spectrally selecting a single harmonic, particularly for low-order harmonics\textemdash in contrast with a sapphire Brewster plate output coupler plus that also requires a grating to separate harmonics\cite{Mills2012}.  The XUV beam (and the corresponding number of grooves illuminated) is relatively small at the GM (3~cm from the focus), which limits the severity of the pulse shearing.  
 We leverage the spatial chirp to achieve a slightly improved energy resolution and spatially filter the XUV with a removable 0.3~mm slit approximately 1.5~m from the source (1~m from the sample).  This slit serves to reduce the spectral bandwidth delivered to the sample, and reduces the spatial chirp. Ultimately, the XUV arrives at the sample location with a size of 0.4 mm x 0.2 mm (FWHM), in the plane normal to the propagation direction.

The fsEC vacuum chamber is normally operated with a pressure of several mTorr, necessitating differential pumping connecting the fsEC XUV source to the ARPES chamber.  This is achieved with three segments of tubing with a diameter of 2~mm placed between the fsEC vacuum chamber and the mirror tank UHV chamber (see Fig.\ref{fig:setup}), with turbo pumps (80~L/s) placed between each of the three segments.  The mirror tank is also pumped with a turbo pump (250~L/s), and a pressure of $<5\times10^{-10}$~Torr is achieved in the mirror tank when the XUV source is operating, leaving the ARPES chamber pressure unaffected by the fsEC gas load.

\section{Equilibrium ARPES: Energy Resolution and Momentum Range}
ARPES is performed in a UHV chamber connected to a Scienta R4000 hemispherical electron analyzer.  The sample is cooled with a custom-built liquid He flow cryostat, capable of reaching temperatures $<4$~K. The vacuum in the main chamber is maintained at $<5\times10^{-11}$~Torr during experiments. The resolution of the analyzer is found to be $<$2~meV in the highest resolution mode, using a helium discharge lamp.  The lamp and laser enter the ARPES chamber from opposite sides, each at an angle of 45~degrees relative to the axis of the analyzer.

\begin{figure*}[t]
{\includegraphics{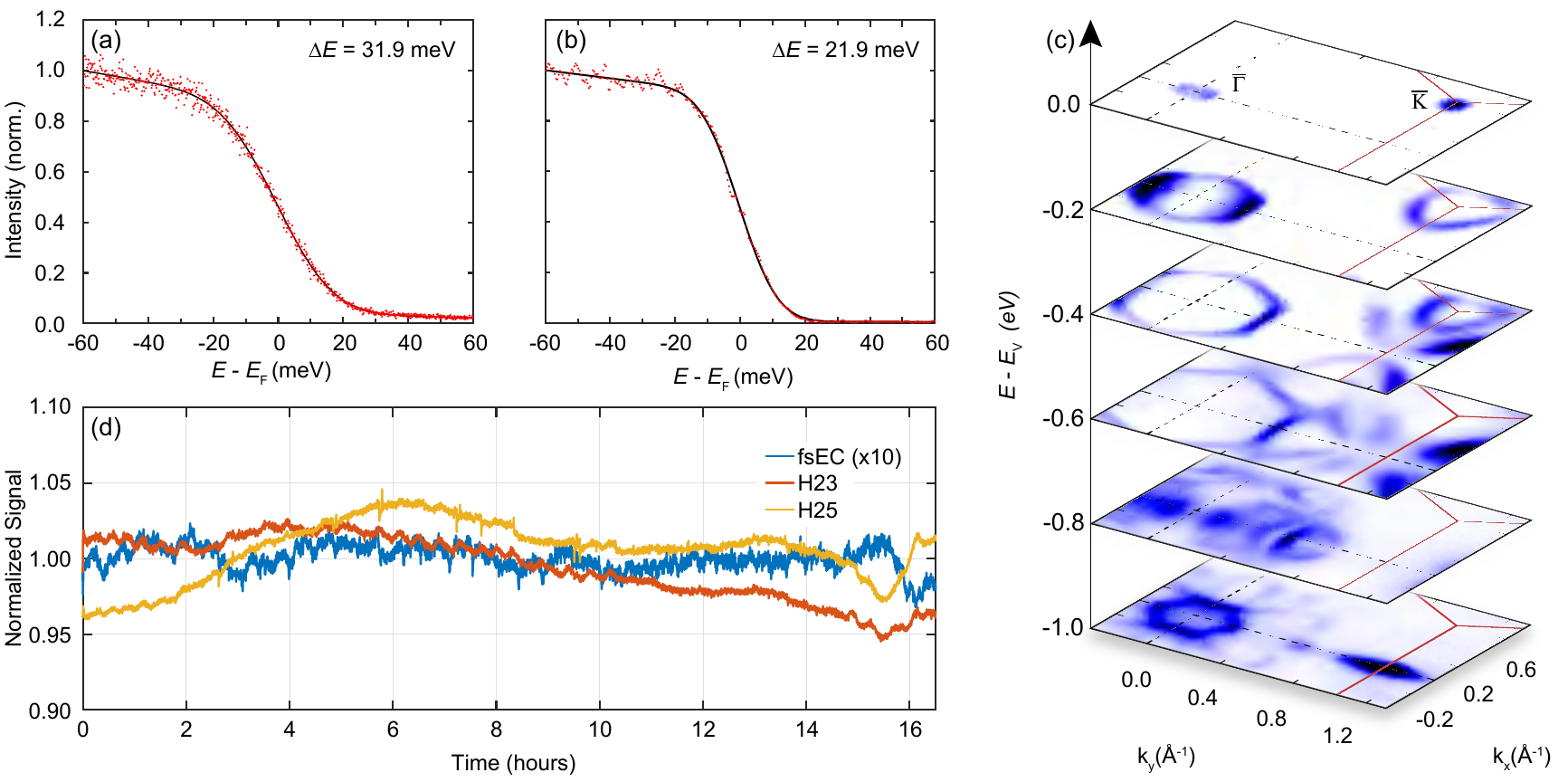}}
\caption{ (a), Unfiltered and  (b), slit-filtered Fermi edge measurements at 4.5 K on freshly evaporated Au used to define the energy resolution ($31.9 \pm0.8$ and $21.9 \pm 0.4)$ meV, respectively. As the analyzer contributes $<1$~meV to the system resolution (determined by an independent measurement using a He lamp) this resolution is essentially the XUV spectral width. (c), ARPES intensity maps of 2H-MoTe$_2$.  The constant binding energy maps at various binding energies, with respect to the valence band maximum, $E_\text{V}$  highlight the momentum reach from $\overline{\Gamma}$ to $\overline{\text{K}}$. (d), A measurement of long-term locking performance demonstrating 0.5\% drift in the fsEC power, and $\pm$5\% drift in H23 and H25 over more than 16 hours of continuous locking.}
\label{fig:char}
\end{figure*}

To evaluate the capabilities of our XUV probe, we characterize the bandwidth, stability, and momentum reach of the H21 (25~eV) in ARPES, as shown in Fig.\,\ref{fig:char}. The energy broadening of the Fermi edge of polycrystalline Au gives a direct measurement of the system energy resolution. In Fig.\,\ref{fig:char}(a) we display the photoemission intensity near the Fermi energy measured at T~=~4~K.  The data were fit with a Fermi-Dirac function convolved with a Gaussian. The FWHM of the Gaussian function defines the overall instrumental energy resolution, and was found to be $\Delta E= 31.9\pm0.8$~meV, shown in Fig.\,\ref{fig:char}(a). Under these conditions, the contribution to the resolution by the electron analyzer is less than the 95\% confidence interval of the fit, thus this resolution is close to the bandwidth of the XUV source.  Inserting the $0.3$~mm slit in the mirror tank chamber reduces the spectral width of the XUV and improves the energy resolution to $\Delta E= 21.9\pm 0.4$~meV, shown in Fig.\,\ref{fig:char}(b), which only reduces the  photoemission count rate by about 25\%.

\begin{figure*}[t]
\includegraphics{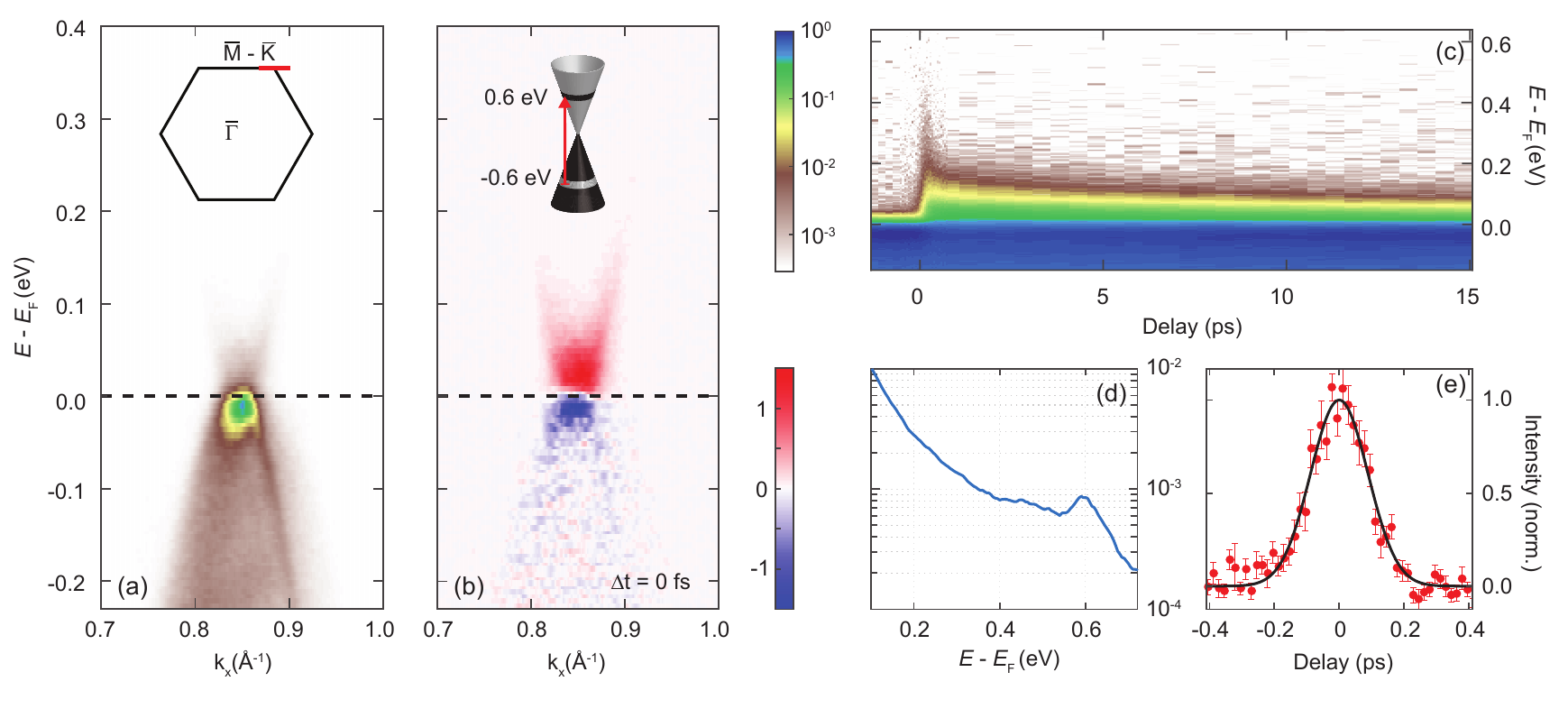}
\caption{(a), Pumped ARPES and (b), differential map of  graphite at zero pump-probe delay with $\overline{\text{M}}\textendash\overline{\text{K}}$ direction parallel to the analyzer slit. Significant excitation of electrons is induced by the pump, and the population (depletion) is clearly visible above (below) $E_F$. (c), Momentum-integrated energy distribution curves (EDC) plotted as a function of pump-probe delay (log color scale, normalized to the signal maximum near $E_F$) illustrating the fast thermalization and subsequent slow recombination of excited electrons, corresponding to electrons excited from -0.6~eV below $E_F$ (d), Momentum-integrated EDC at zero delay (normalized to the signal at $E_F$) showing a distinct peak at 0.6~eV. (e), Cross-correlation (190 fs, FWHM)  of XUV probe and near-infrared (NIR) pump pulses determined from the temporal evolution of the direct population shown in (d).}
\label{fig:GraphiteTRARPES}
\end{figure*}

The benefits of a narrow-bandwidth, high-photon energy probe is seen in the equilibrium band mapping of molybdenum ditelluride (MoTe$_2$). The spins of MoTe$_2$ are locked to the valley degree of freedom at $\overline{\text{K}}$ and $\overline{\text{K}}$$^{\prime}$, making it a candidate material for valleytronic applications \cite{Xiao2012a}. With an in-plane unit cell size of $a=3.492 \mathrm{\AA}$ \cite{Agarwal1972}, the $\overline{\text{K}}$ point is located far from the zone center (1.2~$\mathrm{\AA}^{-1}$) and photoemission of an electron from $\overline{\text{K}}$ effectively requires a minimum photon energy $\ge12$~eV (assuming a 45 degree emission angle). Consequently, an HHG source such as that presented here is essential for time-resolved studies of transition metal dichalcogenides \cite{Crepaldi2017a} and other materials with small lattice constants, such as graphene.  In Fig.\,\ref{fig:char}(c), we demonstrate the momentum-reach of the 25~eV probe over a region of the 2H\textendash MoTe$_2$ Brillouin zone, simultaneously mapping the dispersion of the valence bands at $\overline{\Gamma}$ and $\overline{\text{K}}$.


\section{TR-ARPES on Graphite}
TR-ARPES requires the careful alignment of the optical system to achieve temporal and spatial overlap of the pump and probe beams on the sample.  Spatial overlap is roughly established by imaging the pump and probe beams on a Ce:YAG scintillator crystal, and temporal overlap is achieved with the use of spatial spectral interferometry (SSI) between the pump beam and the residual infrared light (about 50~nW) that scatters from the fsEC GM and travels in the path of the probe beam.  We couple light from these two paths into single-mode optical fiber placed after the ARPES vacuum chamber, which assures spatial mode overlap and greatly improves the SSI signal.   With the pump-probe delay set to zero, we optimize the spatial overlap in real time using graphite by maximizing the photo-induced population detected above $E_F$.

We characterize the temporal properties of our system by tracking electron dynamics in graphite.  Graphite is comprised of layers of carbon atoms arranged in a hexagonal lattice with an in-plane unit cell spacing of  a = 2.46~$\mathrm{\AA}$. Within 1~eV of the Fermi level, its electronic structure consists of three-dimensional valence and conduction bands\textemdash reminiscent of the Dirac cone in graphene\textemdash that meet at the Fermi level and are centered at the corners of the surface-projected hexagonal Brillouin zone. As such, a probe with high photon energy is required to access the states near the Fermi energy with photoemission. We excite a high quality, naturally occurring single-crystal graphite sample (p-doped about 20~meV)  using s-polarized, 1.19~eV pump pulses (120~fs duration). The band structure significantly limits the optical transitions available to the pump. In particular, we expect to observe a direct transition from 0.6~eV  below $E_F$ to 0.6~eV above $E_F$, shown in the inset of Fig.\,\ref{fig:GraphiteTRARPES}(b).

The ARPES map and differential map (obtained by subtracting the equilibrium map from its pumped counterpart) of graphite at zero pump-probe delay along the $\overline{\text{M}}\textendash\overline{\text{K}}$ direction are shown in Fig.\,\ref{fig:GraphiteTRARPES}(a) and (b). Although photoemission intensity is strongest along the $\overline{\Gamma}\textendash\overline{\text{K}}$ direction, the polarization dependent matrix elements allow photoemission from only one branch of the cone, while the $\overline{\text{M}}\textendash\overline{\text{K}}$ cut shows excitation in both branches. With an incidence pump fluence of 18.4~$ \mu\text{J}/\text{cm}^2$ we observe a strong population (depletion) above (below) the Fermi level. The sharp spectral features shown here were acquired in under 10 minutes, demonstrating the ability to resolve the unoccupied band structure with high signal-to-background ratio even in a low pump fluence regime.

Figure\,\ref{fig:GraphiteTRARPES}(c) represents the intensity of the signal in Fig.\,\ref{fig:GraphiteTRARPES}(a), integrated over a range of momentum in k$_x$ of $\pm 0.2\mathrm{\AA}^{-1}$ about $\overline{\text{K}}$ as a function of the pump-probe delay. Near zero delay, we observe a transient signal up to 0.6~eV above $E_F$. This excitation quickly relaxes toward $E_F$ (within 0.5~ps), and the system returns to equilibrium on a timescale  $>$10~ps. The energy resolution of our system allows us to observe the nonthermal optically-excited electron population at 0.6~eV, which appears as a well-defined peak, shown in Fig.\,\ref{fig:GraphiteTRARPES}(d). Both the high energy resolution of our system and the low fluence (enabled by the high repetition rate) make this small signal quite prominent, despite being three orders of magnitude below the signal at $E_F$.  It is in the vicinity of this peak that direct observation of discrete-energy scattering processes like electron-phonon scattering is possible\cite{Na2018}, provided the thermal component of the signal is sufficiently small.   We note that somewhat similar direct-population features have been observed previously in experiments with shorter pulses (and lower energy resolution) and higher pump fluences\cite{Stange2015,Bauer2018}, resulting in large thermal background and  broader spectral features that obscure the signatures of electron-phonon scattering.

The temporal resolution of the system is characterized by the dynamics of the non-thermal direct population of states ({\it i.e.} optically excited electrons) near 0.6 eV above $E_F$. We integrate the angle-integrated energy distribution curves (EDC) within a bandwidth of 125~meV centered at 0.6~eV above $E_F$, and plot this intensity as a function of pump-probe delay in Fig.\,\ref{fig:GraphiteTRARPES}(e). The signal is symmetric with respect to zero delay, which indicates that the scattering processes responsible for decay in this energy range occur on a time scale much shorter than the temporal resolution of our system. A measurement of this electron population as a function of pump-probe delay directly represents the cross-correlation of the pump and probe pulses, and defines the time resolution of the system. A Gaussian fit to the cross-correlation gives an overall time resolution of $190\pm 10$~fs (FWHM). As the pump pulse width is 120~fs, we can infer an XUV pulsewidth of $150\pm11$~fs. Here, the observed XUV pulse duration is largely determined by pulse shearing \cite{WeinerUltrafastOptics} due to broadening by the GM (as described in Section\,\ref{sec:XUVSource}).  In these measurements, the graphite single-crystal domain size is small (approximately 100 $\mu m$), thus we expect the resolving power of the grating ({\it i.e.} the number of grooves illuminated on the GM) to define the observed XUV pulse duration.

\section{Low Pump Fluence Dynamics in Graphite}
\begin{figure}[t]
\includegraphics[width=1\columnwidth]{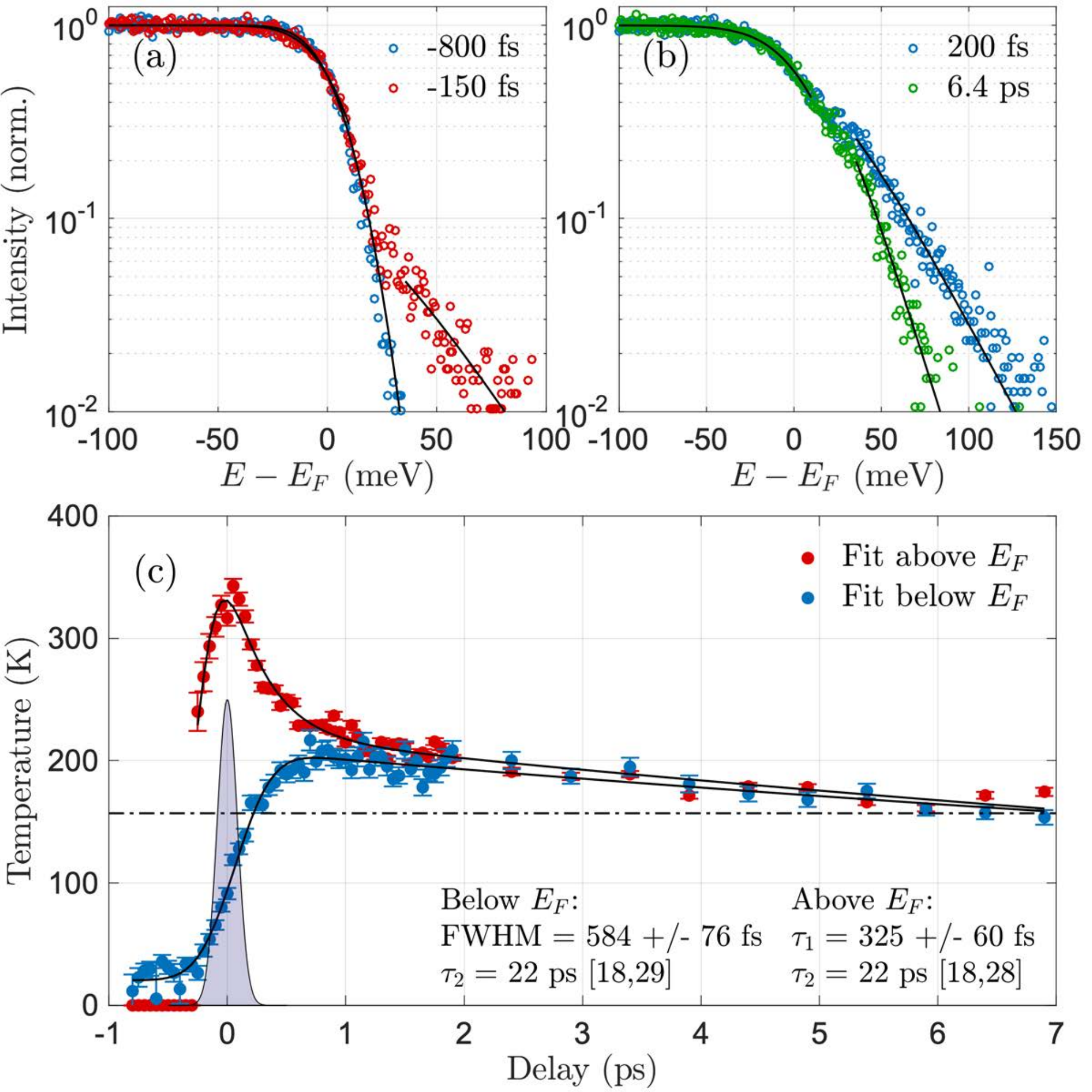}
\caption{Low fluence TR-ARPES data for a pump fluence of 7.7 $\mu$J/cm$^2$ exhibiting different dynamics above and below $E_F$. Panel (a) depicts two energy distribution curves (EDC)  for negative pump-probe delays, including the equilibrium EDC for -800~fs delay and the onset of an observable pumping effect at -150~fs delay ({\it i.e.} before the electron distribution below $E_F$ is significantly modified).  Panel (b) depicts EDCs at different pump-probe delays with three fits to a Fermi-Dirac (FD) distribution.  The fitting procedure in the range -100~meV$<E-E_F<$10~meV yielded the same fit for both delays, while above $E-E_F>$30~meV it produced two different fits. The solid black lines in (a) and (b) represent fits of the EDCs with a FD, where the fit is plotted over the full range of energies for the equilibrium EDC and only over the fitted range of energies for -150~fs, 200~fs, and 6.4~ps delay (the procedure for fitting EDCs above and below $E_F$ is described in the main text). Panel (c) illustrates the effective temperature vs. pump-probe delay as obtained from the FD fits to the EDCs above and below $E_F$, where the error bars represent the 95\% confidence interval on the FD fits to the EDCs.  The solid black lines in (c) represent a fit of the effective temperatures to a Gaussian convolved with a bi-exponential above $E_F$, and a Gaussian convolved with an exponential below $E_F$ (as described in the text).  The parameters resulting from these fits (and corresponding 95\% confidence intervals) are presented in (c).  The dot-dash line in (c) represents the effective temperature fit below $E_F$ for the two EDCs plotted in (b), and the shaded Gaussian near zero delay represents the XUV-NIR cross-correlation from Fig.\,\ref{fig:GraphiteTRARPES}(d).}
\label{fig:Graphite_Te}
\end{figure}

Ultrafast dynamics in graphite and graphene have been studied extensively, and new measurements continue to refine our understanding of these materials \cite{Seibert1990,Kampfrath2005a,Butscher2007,Ishioka2008,Breusing2009a,Scheuch2011b,Ishida2011a,HofmannTRgraphene2013,Chatelain2014,Ulstrup2014a,Johannsen2015,Stange2015,GierzBottleneck2015,Narang2017,Stern2018,Bauer2018}.  During excitation (and shortly afterwards), non-thermal electron distributions can be observed.  Excited carriers thermalize through electron-electron scattering on the sub-50~fs timescale, emission of strongly-coupled optical phonons (SCOP) over about 0.5~ps, and acoustic phonons (AP) on the timescale of many picoseconds.  We note that the majority of previous TR-ARPES studies of graphite and graphene have focused on extremely short timescales using sub-50~fs pulse duration and high pump fluence to identify intrinsic mechanisms of electron-electron scattering.  Relatively few studies have investigated low-energy dynamics around the Fermi level or excitation with low pump fluences. In this section we present our low-fluence study of excited carriers in graphite, in which the high energy resolution of our system allows us to focus on dynamics close to the Fermi level ($|E-E_F|<150$~meV).   While the relaxation dynamics within the conduction band appears instantaneous on the timescale of our experiment, we observe different thermalization dynamics between electron distributions in the conduction and valence bands, which cannot be described by a single Fermi-Dirac (FD) distribution until approximately 1~ps after excitation. 

In particular, we present data corresponding to a measurement time of five hours for a scan of 65 pump-probe delays, at an incident pump fluence of 7.7~$\mu$J/cm$^2$. The sample is naturally occurring, single-crystal graphite p-doped at about 20~meV, as described in the previous section.  In Fig.\,\ref{fig:Graphite_Te}(a) we plot EDCs at early timescales: one showing the equilibrium distribution (-800~fs, blue markers), and a second at -150~fs (red markers) during the initial stages of excitation, {\it i.e.} when the unoccupied band first shows signs of electron occupation above $E_F$ but the distribution below $E_F$ remains largely unaffected (note that data are plotted on a logarithmic scale).  For positive pump-probe delays, shown in Fig.\,\ref{fig:Graphite_Te}(b), we note that EDCs at 200~fs (shortly after excitation, blue markers) and 6~ps (green markers) are fit to the same FD ({\it i.e.} the same temperature) for negative binding energies ($E<E_F$), but they are markedly different for positive binding energies ($E>E_F$). A quantitative description of the measured EDCs requires proper treatment of the density of states and the photoemission matrix elements, which is beyond the scope of this work. Nevertheless, there exist pairs of pump-probe delays that yield identical fits to a FD below $E_F$ and drastically different fits above $E_F$.  This observation is inconsistent with a single temperature description of the electron distributions in the valence and conduction bands.

In an effort to gain new insights into these dynamics, we find that the distribution of electrons above and below $E_F$ can be fit with two FD distributions sharing the same chemical potential.  Accordingly, we describe the non-equilibrium evolution of these carriers in terms of different effective temperatures for each band (valence and conduction), similar to the procedure of Gierz {\it et al.}\cite{Gierz2013}  Specifically, at each pump-probe delay, we fit the measured EDC to a FD convolved with a Gaussian function representing the energy resolution of our system.  The fitting parameters include only the temperature of the distribution and the chemical potential.  In the valence band we restrict the fit to the range 100~meV below $E_F$ to 10~meV above $E_F$ (sample is p-doped by about 20~meV), and in the conduction band, we restrict the fit to a range of 30 to 150 meV above $E_F$, and we constrain the chemical potential to that returned by the fit in the valence band.  By restricting the range of fitting to energies away from the crossing point of the conduction and valence bands, we minimize the effect of the vanishing density of states. The effective temperatures determined by this fitting procedure are shown in Fig.\,\ref{fig:Graphite_Te}(c).

While the effective temperature in the valence band  shows a build-up time on the order of 1~ps, the effective temperature in the conduction band increases rapidly and reaches its maximum at zero pump-probe delay.  The two distributions remain at distinctly different temperatures until approximately 1~ps, after which the temperature dynamics are identical as the system slowly cools.  In the valence band, we find that the effective temperature fits well to a Gaussian convolved with an exponential decay, yielding a 0.58~ps FWHM  for the Gaussian and a 22~ps decay for the exponential. In the conduction band, we convolve our time resolution with a bi-exponential decay, which yields decay time constants of 325~fs and 22~ps.  The observed bi-exponential relaxation for carriers in the conduction band agrees well with previous studies where the decay process is described in terms of a combined action of SCOPs and APs, and the decay times we extract are in agreement with the result of Stange {\it et\,al.}\cite{Stange2015}.  However, contrary to Stange {\it et\,al.}, where only one FD was used for describing the electron dynamics, our experimental evidence suggests that optically injected electrons form a hot electronic distribution separate from the depleted valence band, and that this excess stored thermal energy is released to the valence electronic distribution over a time scale consistent with the emission of SCOPs. 

These results demonstrate that TR-ARPES performed at a high repetition rate with high energy resolution allows one to track the dynamics of excited carriers at pump fluence levels that leave the system temperature relatively low through the excitation process. This is promising for future studies of sensitive quantum matter systems, such as high-Tc superconductors, in which one might wish to excite the material without driving the system too far from its superconducting transition temperature. Furthermore, we emphasize that in order to gain a complete understanding of dynamics in material systems, high time resolution measurements must complement high energy resolution measurements, and to characterize specific dynamics within some region of a material's band structure or a specific scattering process ({\it e.g.} electron-phonon), the time and energy resolutions must be balanced appropriately.

\section{Conclusion}
We have developed an XUV laser source for TR-ARPES delivering in excess of 10$^{11}$~photons/s to a sample, with a tuning range of 8-40~eV and a repetition rate of 60~MHz.  We presented a detailed characterization of the source at 25~eV demonstrating a time resolution of 190~fs, and an energy resolution of 22~meV.  Our source breaks new and significant ground for studies of the electronic structure of complex materials, within the perturbative pump excitation regime over the full Brillouin zone. This suggests the intriguing opportunity to explore and carefully resolve the dynamics of low-energy spectroscopic features of quantum matter, such as the superconducting \cite{Boschini2018, Cilento2018} or charge-ordered many-body gaps \cite{Rohwer2011}. In addition, the future development of novel pump excitation schemes will allow for a full characterization of out-of-equilibrium phenomena and raises the opportunity for studies of multi-color quantum coherent control over the entire Brillouin zone \cite{Stevens2003}.

\section{Acknowledgements}
This research was undertaken thanks in part to funding from the Max Planck-UBC-UTokyo Centre for Quantum Materials and the Canada First Research Excellence Fund, Quantum Materials and Future Technologies Program. The work at UBC was supported by the Gordon and Betty Moore Foundation's EPiQS Initiative, grant GBMF4779, the Killam, Alfred P. Sloan and Natural Sciences and Engineering Research Council of Canada's (NSERC's) Steacie Memorial Fellowships (A.D.), the Alexander von Humboldt Fellowship (A.D.), the Canada Research Chairs Program (A.D.), NSERC, Canada Foundation for Innovation (CFI), British Columbia Knowledge and Development Fund (BCKDF), and CIFAR Quantum Materials. E. Razzoli acknowledges support from the Swiss National Science Foundation (SNSF) grant no. P300P2\_ 164649. S. Zhdanovich acknowledges support from NSERC, 454770-2014-PDF. The authors would like to thank Thomas K. Allison for many insightful discussions, and Emma Fajeau for implementing the autolock hardware and helping to optimize its performance.

\bibliographystyle{apsrev4-1}

\end{document}